\documentclass{iopart}

\usepackage{iopams} 
\usepackage{graphicx}
\usepackage{epstopdf} 

\def\ket#1{\mathinner{|{#1}\rangle}}
\newcommand{\braket}[2]{\langle #1|#2\rangle}

\newcommand{\expectation}[1]{\langle #1\rangle}
\usepackage{setstack}
\begin{document}

\title[Continuous Formation of Vibronic Ground State RbCs Molecules via PA]{Continuous Formation of Vibronic Ground State RbCs Molecules via Photoassociation}

\author{C.D. Bruzewicz, Mattias Gustavsson, Toshihiko Shimasaki and D. DeMille}

\address{Department of Physics, Yale University, 217 Prospect St. New Haven, CT 06511 USA}
\ead{colin.bruzewicz@yale.edu}
\begin{abstract}
We demonstrate the direct formation of vibronic ground state RbCs molecules by photoassociation of ultracold atoms followed by radiative stabilization. The photoassociation proceeds through deeply-bound levels of the (2)$^{3}\Pi_{0^{+}}$ state. From analysis of the relevant free-to-bound and bound-to-bound Franck-Condon factors, we have predicted and experimentally verified a set of photoassociation resonances that lead to efficient creation of molecules in the $v$=0 vibrational level of the X$^{1}\Sigma^{+}$ electronic ground state.  We also compare the observed and calculated laser intensity required to saturate the photoassociation rate.  We discuss the prospects for using short-range photoassociation to create and accumulate samples of ultracold polar molecules in their rovibronic ground state.
\end{abstract}

\pacs{33.20.Tp, 34.80.Gs, 37.10.Mn, 82.30.Mn}
\submitto{\NJP}
\maketitle
\section{Introduction}
Rapid advances in the field of ultracold molecular physics have motivated the development of new methods to create large trapped samples of polar molecules \cite{carr2009cold,quéméner2012ultracold}. Using methods such as photoassociation (PA) and Feshbach association, molecules can be formed from pre-cooled atomic samples, inheriting the ultracold translational temperature of the precursor atoms.  Generally the molecules created with these techniques are weakly bound and thus not significantly polar \cite{aymar:204302}. Further, they are prone to inelastic scattering and must be quickly transferred to their absolute ground state to avoid rapid loss \cite{PhysRevLett.96.023202,Ospelkaus12022010,hudsoninelastic2008,deiglmayr2011inelastic}. Transfer from weakly-bound levels to the vibronic ground state has been demonstrated using two-photon processes such as stimulated emission pumping and stimulated Raman adiabatic passage \cite{winklercoherent2007,nihigh2008,sageoptical2005,danzlquantum20081,PhysRevLett.105.203001}. These transfer processes add significant experimental complexity, and suffer from the fact that they can only be applied once per experimental realization, limiting the ultimate number of ground state molecules.

A different method has also been demonstrated, where molecules are photoassociated to deeply-bound, electronically-excited levels. Once formed, these short-range states can spontaneously decay to deeply-bound levels in the X$^{1}\Sigma^{+}$ potential. However, the cross-sections for photoassociation to these states, where the internuclear distances in the molecule are short, are generally expected to be much smaller than for long-range, weakly-bound levels. This method has been used to create deeply-bound K$_{2}$, NaCs, and KRb molecules; in the case of LiCs, rovibronic ground state X($v$=$J$=0) molecules have been created, albeit at an anomalously lower rate than for X($v$=0,$J$=2) molecules \cite{PhysRevLett.84.246,PhysRevA.84.061401,PhysRevA.86.053428,deiglmayrformation2008}. Due to the dissipative nature of this process (from the spontaneous decay step), molecular formation can proceed continuously. Thus in the presence of an optical trap, it may be possible to accumulate ground state molecules as they are continuously produced. RbCs is one of an especially interesting class of molecules to consider forming with this method. The 2 RbCs $\to$ Rb$_{2}$ + Cs$_{2}$ reaction is endothermic in the rovibronic ground state, making a sample of X($v$=$J$=0) RbCs molecules collisionally stable \cite{PhysRevA.81.060703}. Hence, accumulation of a large sample of rovibronic ground state RbCs may be possible, without quickly losing them due to inelastic scattering. 

Recently a set of short-range PA resonances in RbCs was reported in Refs. \cite{C1CP21497G,PhysRevA.85.013401,bouloufa2012formation,PhysRevA.87.054701}. The rotational constants correspond to an internuclear separation comparable to that at the bottom of the X$^{1}\Sigma^{+}$ potential well, thus making it possible that these states could have large Franck-Condon factors (FCFs) for decay to the X($v$=0) state.  In Refs. \cite{C1CP21497G,PhysRevA.85.013401,bouloufa2012formation,PhysRevA.87.054701}, it was shown that levels in the shallow a$^{3}\Sigma^{+}$ state were being populated, but it was not determined if the X$^{1}\Sigma^{+}$ state was populated as well.  One state in this set was studied further in Ref. \cite{PhysRevA.85.013401}, where the state's electric dipole moment and the saturation intensity of PA were measured. Ref. \cite{bouloufa2012formation} gives theoretical analysis of possible decay pathways for this particular state.

Here we report population of the X$^{1}\Sigma^{+}$($v$=0) level (and other deeply-bound vibrational levels of the state) by PA to a separate but nearby set of levels, followed by spontaneous decay. We show that the PA resonances we observe correspond to the $\Omega$=$0^{+}$ component of the (2)$^{3}\Pi_{0}$ state, and we also assign the levels studied in Refs. \cite{C1CP21497G,PhysRevA.85.013401,bouloufa2012formation,PhysRevA.87.054701} to the $\Omega$=0$^{-}$ component of the same state. We also report measurements of the PA rate and its dependence on laser intensity, and compare our measurements to calculated values in order to predict conditions necessary to maximize the rate of formation of X($v$=0) RbCs molecules. As we outline below, this set of results shows a pathway to the formation and accumulation of large samples of trapped, ultracold X$^{1}\Sigma^{+}$($v$=$J$=0) RbCs molecules. 

\begin{figure}
\begin{center}
\includegraphics[width=0.6\columnwidth]{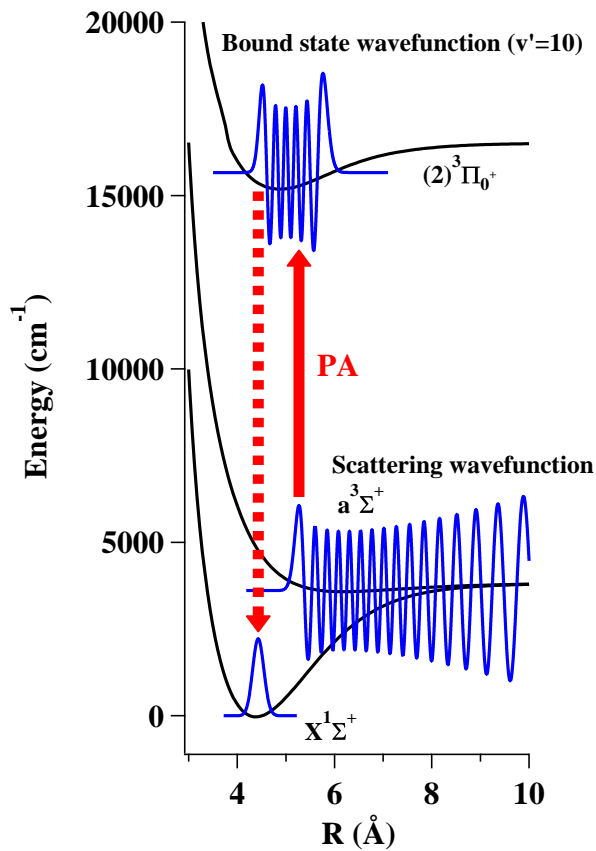}
\end{center}
\caption{Calculated wavefunctions for free-to-bound photoassociation and bound-to-bound [(2)$^{3}\Pi_{0^{+}}$($v'$=10) $\to$ X$^{1}\Sigma^{+}$($v$=0)] spontaneous decay. Wavefunction amplitudes are not drawn to scale.  Potential energy curves are from values in Ref. \cite{0953-4075-35-6-307}.}
\label{rkr}
\end{figure}

\section{Choosing a Pathway to X$^{1}\Sigma^{+}$ ($v$=0)}
Stimulated by the results of Refs. \cite{C1CP21497G,PhysRevA.85.013401,bouloufa2012formation,PhysRevA.87.054701} demonstrating short-range PA in RbCs, we sought to find a PA state that would create X$^{1}\Sigma^{+}$($v$=0) most efficiently.  The requirements for such a state are that its electronic state must satisfy selection rules for transitions from the initial atomic scattering state and to the final X $^{1}\Sigma^{+}$ state, and that it must have substantial FCFs both for the free-to-bound PA transition and the bound-to-bound spontaneous decay to X($v$=0). Detailed spectroscopic data from Ref. \cite{doi:10.1021/jp803360w}, obtained by observing transitions from the X($v$=0) state in a cold molecular beam, showed that the (2)$^{3}\Pi_{0^{+}}$ state satisfies the requirements for coupling to the X($v$=0) state in RbCs. The (2)$^{3}\Pi_{0^{+}}$ state also has the desirable features of simple rotational and hyperfine structure. To determine which vibrational levels in the (2)$^{3}\Pi_{0^{+}}$ state should lead to the most efficient formation of $v$=0 molecules in the X$^{1}\Sigma^{+}$ state, we analyzed the FCFs for both the free-to-bound photoassociation and bound-to-bound spontaneous decay steps of the formation process. 

\begin{figure}
\begin{center}
\includegraphics[width=\columnwidth]{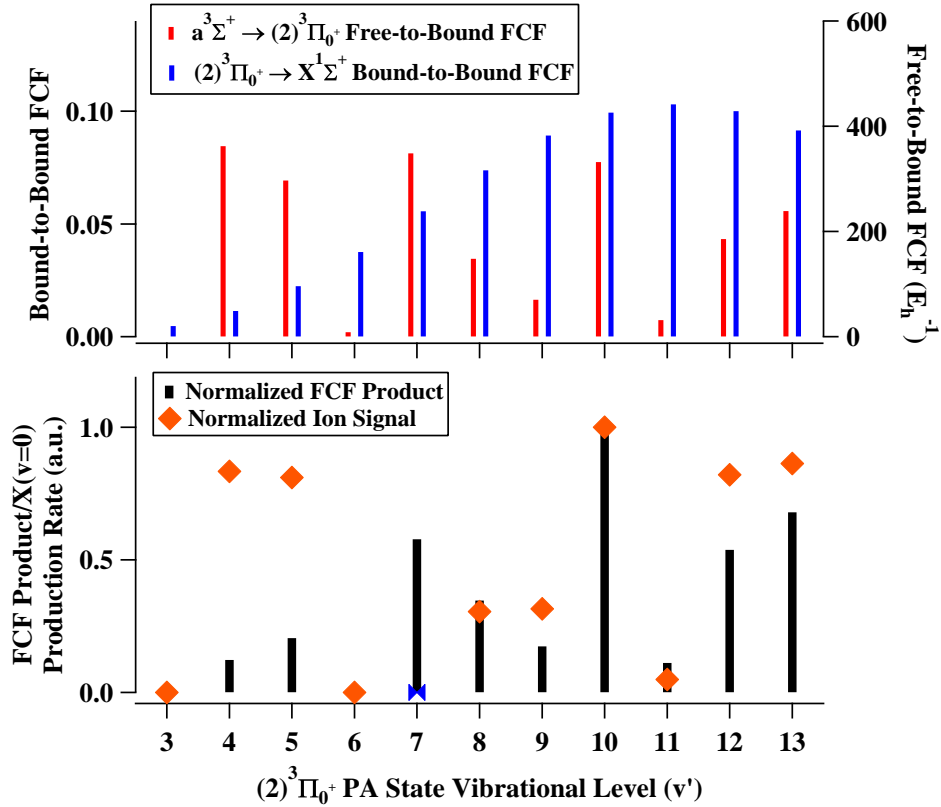}
\end{center}
\caption{Top: Calculated Franck-Condon factors for different vibrational levels, $v'$, in the (2)$^{3}\Pi_{0^{+}}$ state. The bound-to-bound FCFs (blue) are evaluated for transitions to the X$^{1}\Sigma^{+}$($v$=0) state. Free-to-bound FCFs (red) are calculated for an atomic scattering wavefunction at an energy corresponding to a temperature of 100 $\mu$K.  Bottom: Calculated products of free-to-bound and bound-to-bound FCFs are shown in black. Measured X$^{1}\Sigma^{+}$($v$=0) ion signals are shown as points. Both are scaled such that the height of $v'$=10 is 1. Blue point indicates a level ($v'$=7) for which the PA laser itself strongly affects the Cs density (see Section 4.1).}
\label{FCF}
\end{figure}

We verified the bound-to-bound FCFs calculated in Ref. \cite{doi:10.1021/jp803360w} using the LEVEL 8.0 program \cite{leroylevel}, with the (2)$^{3}\Pi_{0^{+}}$ level positions from Ref. \cite{doi:10.1021/jp803360w} and the X$^{1}\Sigma^{+}$ molecular potential from Ref. \cite{PhysRevA.83.052519}. In order to determine the free-to-bound FCFs, we solved the Schr\"{o}dinger equation numerically for the atomic scattering state in the presence of the a$^{3}\Sigma^{+}$ potential characterized in Ref. \cite{PhysRevA.83.052519}. The resulting wavefunctions, shown in Figure \ref{rkr}, were normalized by scaling the entire wavefunction such that the amplitude at large internuclear separation matched the result from a long-range wavefunction calculated in the WKB approximation, with the standard energy normalization \cite{landau1977quantum}. From the product of the two FCFs, shown in Figure \ref{FCF}, we predicted X$^{1}\Sigma^{+}$($v$=0) formation rates through numerous vibrational levels of the (2)$^{3}\Pi_{0^{+}}$ state. Based on these results, we predicted that the (2)$^{3}\Pi_{0^{+}}$($v'$=10) level (PA energy=11817.16 cm$^{-1}$) has the large free-to-bound and bound-to-bound FCFs necessary to efficiently create X$^{1}\Sigma^{+}$($v$=0) state molecules.  This analysis assumes no enhancement of the PA rate due to resonant coupling in the excited PA state and should therefore be widely applicable to other levels and molecules.

\section{Experimental}
The experimental setup is very similar to work described earlier in Refs. \cite{sageoptical2005,hudsoninelastic2008}.  $^{133}$Cs and $^{85}$Rb atoms, dispensed from heated getter sources, are trapped in two independent overlapped spatial forced dark-SPOT MOTs with densities (atom numbers) of $n_{\mathrm{Rb}}\approx3\times$10$^{10}$ cm$^{-3}$ ($N_{\mathrm{Rb}}\approx5\times10^{6}$) and $n_{\mathrm{Cs}}\approx5\times$10$^{10}$ cm$^{-3}$ ($N_{\mathrm{Cs}}\approx10^{7}$). The atomic temperatures were measured by time-of-flight expansion as $T_{\mathrm{Rb}}\approx$80 $\mu$K and $T_{\mathrm{Cs}}\approx$100 $\mu$K.  

To drive the PA transitions, we continuously illuminate the atomic clouds with $\sim$250 mW of narrowband ($\sim$1 MHz) light from a Ti:Sapphire laser with a beam waist (1/$e^{2}$ power radius) of 100 $\mu$m. Atoms in their lower hyperfine ground states ($F_{\mathrm{Cs}}$=3, $F_{\mathrm{Rb}}$=2) collide in the presence of a resonant photon and are promoted to a selected rovibrational level in an electronically-excited molecular state. These molecules quickly decay to numerous vibrational and rotational levels associated with the electronic ground states.

To state-selectively detect the electronic ground state molecules, we use resonance-enhanced multiphoton ionization (REMPI) spectroscopy. A tunable pulsed dye laser (DCM dissolved in DMSO) operating at 10 Hz near 650 nm (pulse energy $\approx$300 $\mu$J, $d\approx$4 mm) resonantly excites a single vibrational level, whose population is subsequently photoionized by a second laser pulse at 532 nm (pulse energy $\approx$2 mJ, $d\approx$4 mm) that arrives approximately 10 ns later. A continuously applied voltage of 2 kV accelerates the ions to a monolithic Channeltron ion detector. The signal from the Channeltron is sent through a transimpedance amplifier and then to DAQ electronics. The signal is analyzed by time-of-flight mass spectroscopy to differentiate molecular RbCs$^{+}$ ions from other atomic and molecular species. The molecular ion signal is calculated by integrating the time-of-flight trace over the appropriate narrow time window corresponding to the arrival of RbCs$^{+}$.  The spectral resolution of this detection scheme is limited by the relatively broad resonant dye laser pulse ($\sim$0.3 cm$^{-1}$). The relevant vibrational levels are spaced by more than 10 cm$^{-1}$ and are well-resolved. The rotational and hyperfine splittings, however, are not resolved by the pulsed laser.

\section{Results and Analysis}
\subsection{Photoassociation Spectroscopy of \upshape{(2)}$^{3}\Pi_{0^{+}}$}
We have observed several deeply-bound PA lines correlated to vibrational levels in the $(2)^{3}\Pi_{0^{+}}$ state, including $v'$=1,2,4,5,8-13. The expected resonance frequencies for these lines were calculated by subtracting the energy difference between the bottom of the X$^{1}\Sigma^{+}$ potential well and the Rb 5$s_{1/2}$ ($F$=2) + Cs 6$s_{1/2}$ ($F$=3) atomic asymptote (3836.141 cm$^{-1}$) \cite{PhysRevA.83.052519} from the (2)$^3\Pi_{0^{+}}$ term energies given in Ref. \cite{doi:10.1021/jp803360w}. The experimentally observed line positions and rotational constants were found to be in excellent agreement with these calculated values, with deviations at a level consistent with the accuracy of our wavelength determination ($\sim$0.007 cm$^{-1}$ for total energy and $\sim$0.0002 cm$^{-1}$ for rotational constant). We failed to observe PA to the $v'$=3,6,7 levels; this is consistent with the very low calculated free-to-bound Franck-Condon factors for $v'$=3,6. For $v'$=7, the PA excitation frequency is very close to an atomic Cs resonance;  our intense PA laser perturbs the MOT here, precluding efficient PA to this level. 

\begin{figure}
\begin{center}
\includegraphics[width=0.85\columnwidth]{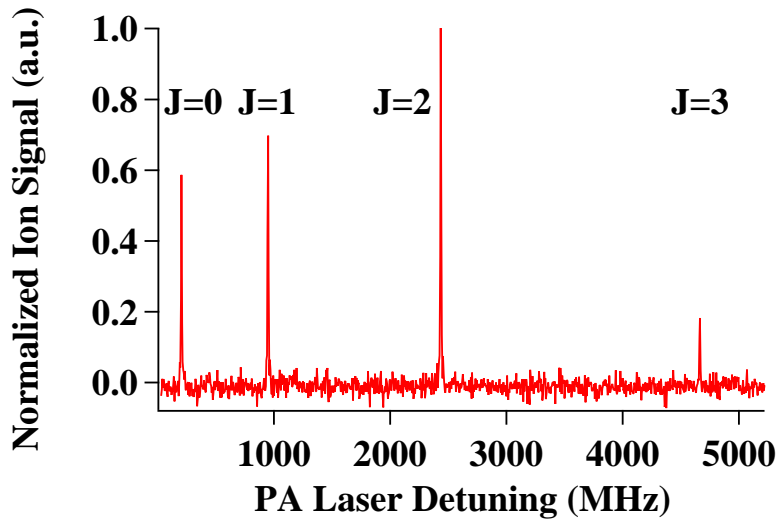}
\end{center}
\caption{\label{parot}PA spectrum showing rotational structure of the (2)$^{3}\Pi_{0^{+}}$($v'$=10) level. The measured rotational constant, (B$_{v'=10}$=0.0124(2) cm$^{-1}$, is consistent with the value given in Ref. \cite{doi:10.1021/jp803360w} for this level (B$_{v'=10}$=0.0123 cm$^{-1}$). Relative peak areas are consistent with calculated results using the formalism in Ref. \cite{calcpaper}. Peak areas (rather than heights) were used to account for observed broadening due to the Stark effect from the applied electric field used to accelerate the molecular ions. The $J$=4 rotational level has not been observed at the predicted location, which is consistent with excitation of $\ell\leq2$ scattering states to an $\Omega=0^{+}$ PA state.}
\end{figure}

Recently Ji \textit{et al.} \cite{PhysRevA.85.013401} and Fioretti and Gabbanini \cite{PhysRevA.87.054701} observed different nearby short-range PA lines that they assigned to the (2)$^{3}\Pi_{0^{+}}$ state. We believe that assignment is incorrect, and that instead the levels they observed must belong to the $\Omega$=0$^{-}$ component of the (2)$^{3}\Pi_{0}$ state. Our conclusion is based primarily on the very precise agreement of our observations with the data in Ref. \cite{doi:10.1021/jp803360w}. By contrast, the assignment in Refs. \cite{PhysRevA.85.013401,PhysRevA.87.054701} relied on much less precise comparison to ab initio calculations as a way to distinguish between assignment to $\Omega$=0$^{+}$ versus $\Omega$=0$^{-}$. To further validate our conclusion about the state assignment, we calculated the rotational line strengths for PA to deeply-bound levels in both the $\Omega$=0$^{+}$ and the $\Omega$=0$^{-}$ components of a $^{3}\Pi_{0}$ state; the detailed calculation will be given in a separate paper \cite{calcpaper}.  Based on this analysis, we predict strikingly different rotational line strength distributions for $\Omega$=0$^{+}$ versus $\Omega$=0$^{-}$ states. In particular, for the $\Omega$=0$^{+}$ state we find that the ratio of signal sizes for $J$=2 to that for $J$=0 should be 2, which agrees well with the experimental result of 1.8 from Figure \ref{parot}. Similarly, we find that the $J$=1 to $J$=3 signal size ratio for the $\Omega$=0$^{-}$ state should have the value of 4, which appears to agree very well with the result in Figure 3 in Ref. \cite{PhysRevA.85.013401}. Based on these results, we assign the PA lines found in those earlier works to the nearby (2)$^{3}\Pi_{0^{-}}$ state with high confidence.

\subsection{Determination of  X$^{1}\Sigma^{+}(v)$ Vibrational Populations}
\begin{figure}
\begin{center}
\includegraphics[width=0.8\columnwidth]{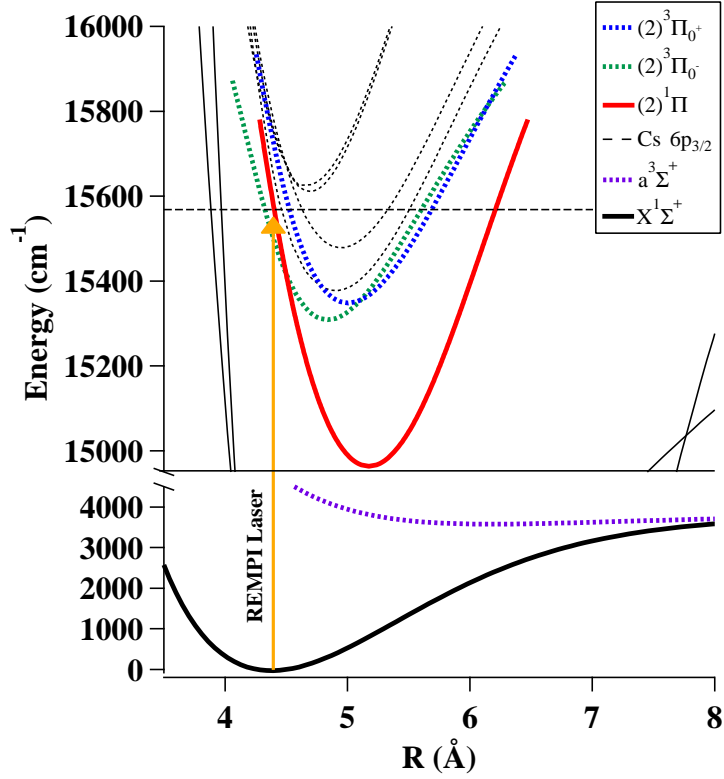}
\end{center}
\caption{\label{rbcscurves} Relevant potential energy curves for RbCs from values in Refs. \cite{PhysRevA.87.054701,0953-4075-35-6-307,doi:10.1021/jp803360w,PhysRevA.83.052519,gustavsson1988spectroscopic}. Photoassociation proceeds through deeply-bound levels of the (2)$^{3}\Pi_{0^{+}}$ state. Spontaneous decay then populates the X$^{1}\Sigma^{+}$ and a$^{3}\Sigma^{+}$ states. The resulting vibrational population in the X$^{1}\Sigma^{+}$ state is measured through REMPI analysis of X$^{1}\Sigma^{+} \to$ (2)$^{1}\Pi$ transitions. Single-photon transitions to the nearby (2)$^{3}\Pi_{0^{-}}$ state are forbidden and do not contribute to the REMPI signal. Thick (thin) lines denote potentials extracted from experimental data (ab initio calculations). Solid (dotted) lines denote singlet (triplet) states. Dashed horizontal line shows the 6$p_{3/2}$ atomic Cs asymptote. Potentials incapable of supporting bound states in the range of internuclear separations of interest have been omitted.}
\end{figure}

\begin{figure}
\begin{center}
\includegraphics[width=\columnwidth]{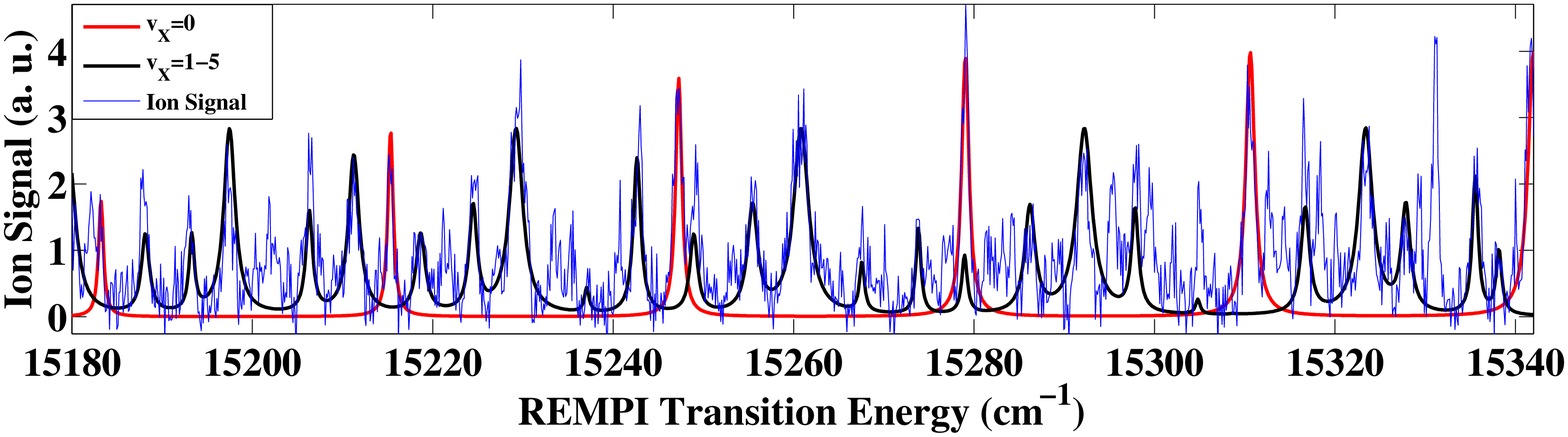}
\end{center}
\caption{\label{REMPI} REMPI scan of vibrational progression of the X$^{1}\Sigma^{+}$($v_{x}$) $\to$ $(2)^{1}\Pi$ ($v'_{\Pi}$) transitions. For this data, the PA laser frequency is locked to the $J$=1 rotational line of the (2)$^{3}\Pi_{0^{+}}$ state (near 11817.18 cm$^{-1}$) while the frequency of the pulsed dye laser is scanned. Red (black) fit line shows the predicted spectra arising from population in X$^{1}\Sigma^{+}$($v_{x}$=0) (sum of $v_{x}$=1-5). Line positions for the fit are based on spectroscopic data in Ref. \cite{doi:10.1021/jp803360w}; peak heights and widths are set by calculated FCFs and fit vibrational populations, $p_{v_{x}}$.}
\end{figure}

Following spontaneous decay from the PA state, numerous rovibrational levels in the electronic ground a$^{3}\Sigma^{+}$ and X$^{1}\Sigma^{+}$ states are populated. To determine which vibrational levels in the X$^{1}\Sigma^{+}$ state were populated, we analyzed 1+1 REMPI spectra with the resonant first step, shown in Figure \ref{rbcscurves}, proceeding through the well-characterized (2)$^{1}\Pi$ state \cite{gustavsson1988spectroscopic,doi:10.1021/jp803360w}.  For select portions of the scanned region, as seen in Figure \ref{REMPI}, the spectra are very simple and allow for straightforward assignment of the observed transitions. We conclude that the $v$=0-5 vibrational levels of the X$^{1}\Sigma^{+}$ state are significantly populated following spontaneous decay of typical PA resonances in our experiment. 

From our REMPI data, we can extract a measure of the population, $p_{v_{x}}$, in a given X$^{1}\Sigma^{+}$ vibrational level, $v_{x}$. The spectra are fit to a sum of Lorentzian curves given by the following form:
\begin{eqnarray}
S(f)=\sum_{v_{x}v'_{\Pi}}\frac{(\Gamma+\beta\cdot F(v_{x},v'_{\Pi}))^{2}(1-e^{-\alpha\cdot F(v_{x},v'_{\Pi})})}{(f-f_{0}(v_{x},v'_{\Pi}))^2+(\Gamma+\beta\cdot F(v_{x},v'_{\Pi}))^{2}}p_{v_{X}}.
\end{eqnarray}
Here $S(f)$ is the signal at laser frequency $f$, $v'_{\Pi}$ is the vibrational state in the (2)$^{1}\Pi$ state, $\Gamma$ is the spectral width of the pulsed dye laser, $f_{0}(v_{x},v'_{\Pi})$ is the transition frequency from $v_{x}$ to $v'_{\Pi}$, $\alpha, \beta$ are fit parameters that are constant for all transitions, $F$ is the known FCF (calculated from the data in Ref. \cite{gustavsson1988spectroscopic}) for the transition at $f_{0}$, and $p_{v_{x}}$ is an adjustable X$^{1}\Sigma^{+}$ vibrational population fit parameter. Here $\alpha$ and $\beta$ describe the saturation and power broadening, respectively, of the laser-driven transitions. The values of $p_{v_{x}}$ are adjusted by hand to fit the data, but are the same for all transitions originating from a single X$^{1}\Sigma^{+}$ vibrational level, $v_{x}$. Results of applying this procedure, for the spectrum arising from PA into the $(2) ^{3}\Pi_{0^{+}}$($v'$=10) level, are shown in Table \ref{fracpop}. 

Additionally, we determined relative values for X$^{1}\Sigma^{+}$($v$=0) molecular production rates for PA through various vibrational levels, $v'$, of the (2)$ ^{3}\Pi_{0^{+}}$ state, summarized in Figure \ref{FCF}. We see generally good qualitative agreement with our calculated rates based on our free-to-bound and bound-to-bound FCF calculations. The primary discrepancies come from an apparent over-production for PA through $v'$=4,5. We believe this may be due to a local increase in the Cs atomic density at the PA laser position, due to dipolar trapping of Cs atoms in the PA laser beam. For these levels, the intense PA laser is approximately 75-100 cm$^{-1}$ red-detuned relative to the Cs D2 line, creating a potential well a few tens of $\mu$K deep for Cs atoms under our conditions, possibly resulting in increased Cs density and hence increased PA rate. Lastly, modest saturation of PA was seen for (2)$^{3}\Pi_{0^{+}}$($v'$=10) at the PA laser intensity ($I_{PA}\!\approx$550 W/cm$^{2}$) used here, and is likely for vibrational levels with similarly high free-to-bound FCFs ($v'$=4,5). This would slightly reduce the observed X($v$=0) production rate, relative to the expectation from the calculated FCF product.
 
\begin{table}
\begin{center}
\begin{tabular}{|c|c|}
\hline
X$^{1}\Sigma^{+}$ Vibrational Level ($v_{x}$) & Relative Population ($p_{v_{x}}/p_{v_{0}}$) \\ \hline
0 & 1 \\ \hline
1 & 0.67 \\ \hline
2 & 0.5 \\ \hline
3 & 0.33 \\ \hline 
4 & 0.33 \\ \hline
5 & 0.17 \\ \hline
6-10 & 0.33 \\
\hline
\end{tabular}
\end{center}
\caption{\label{fracpop} Relative X$^{1}\Sigma^{+}$($v_{x}$) vibrational level populations, $p_{v_{x}}/p_{v_{0}}$, following spontaneous decay from (2)$^{3}\Pi_{0^{+}}$($v'$=10) extracted from REMPI spectroscopy data.}
\end{table}

We also performed similar REMPI analysis of the previously discovered PA line at 11724.08 cm$^{-1}$ first seen in Ref. \cite{C1CP21497G}, which we now assign to the (2)$^{3}\Pi_{0^{-}}$ state. Selection rules forbid direct decay of this $\Omega=0^{-}$ state to the X$^{1}\Sigma^{+}$($\Omega$=0$^{+}$) state. Nevertheless, significant population of the X$^{1}\Sigma^{+}$($v$=0) state occurred, at a rate comparable to that seen for production from the allowed decay of the (2)$^{3}\Pi_{0^{+}}$($v'$=10) PA line. We discuss a possible explanation for this observation later, in Section 4.4.

\subsection{Saturation of Molecular Production}
\begin{figure}
\begin{center}
\includegraphics[width=0.75\columnwidth]{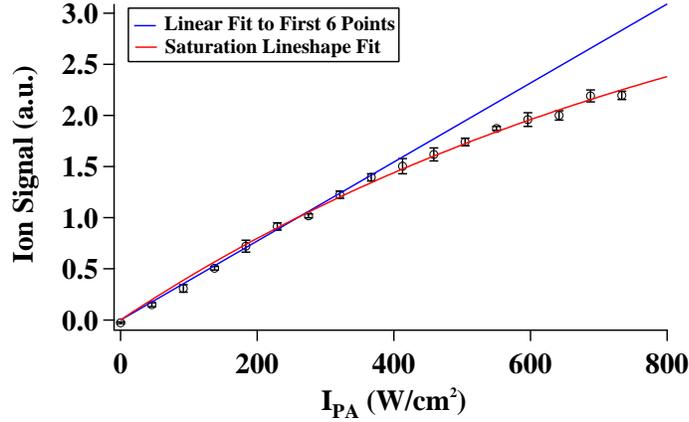}
\end{center}
\caption{\label{saturation} Onset of saturation in molecular production with increased laser intensity. Here the PA laser is locked to the $J$=1 line of the (2)$^{3}\Pi_{0^{+}}$($v'$=10) level. Fit to saturation behavior of Equation (2) (red) allows for extraction of the saturation intensity, $I_{\mathrm{sat}}$.}
\end{figure}

We also investigated the molecular production rate as a function of PA laser intensity. To model the saturation behavior, we recall from Ref. \cite{bohnsemianalytic1999} that the ground state molecular formation rate coefficient, $K$, is proportional to
\begin{equation}
K\propto \frac{I_{\mathrm{PA}}}{(I_{\mathrm{PA}}+I_{\mathrm{sat}})^{2}},
\end{equation}
where $I_{\mathrm{sat}}$ is the PA laser intensity that maximizes ground state molecular formation. By measuring the molecular production as a function of laser intensity, we can extract an approximate value for $I_{\mathrm{sat}}$, as shown in Figure \ref{saturation}. The fit yields the value for I$_{\mathrm{sat}}=$ 4(2) kW/cm$^{2}$, where the uncertainty is dominated by imprecise knowledge of the effective PA laser beam size at the position of the atoms. We can compare this fit value to one estimated using our free-to-bound FCF and a crudely approximated value for the product of angular factors and the electronic matrix element (taken here as $2ea_{0}$, where $e$ is the electron charge and $a_{0}$ is the Bohr radius). With this assumption, we estimate a saturation intensity for PA to the (2)$^{3}\Pi_{0^{+}}$ ($v'$=10) state of I$_{\mathrm{sat}}\sim$13 kW/cm$^{2}$, in crude qualitative agreement with our fit value. 

\subsection{Molecular X$^{1}\Sigma^{+}$($v$=0) Production Rate}
In this section we look to compare observations and calculations of the absolute production rate for X$^{1}\Sigma^{+}$($v$=0) molecules in our experiment. As in Ref. \cite{bohnsemianalytic1999}, the PA rate, $\Gamma_{\mathrm{PA}}$, can be written as
\begin{equation}
\hbar\Gamma_{\mathrm{PA}}=2\pi|\braket{\psi_{\mathrm{PA}}}{\vec{d}\cdot \vec{E}|\chi(R)}|^{2},
\end{equation}
where $\ket{\psi_{\mathrm{PA}}}$ is the PA state wavefunction, $\ket{\chi(R)}$ is the energy-normalized atomic scattering state wavefunction, $\vec{d}$ is the dipole operator, and $\vec{E}$ is the electric field from the PA laser. Using the Born-Oppenheimer approximation, we can factor the total matrix element into electronic and nuclear components and approximate the PA rate as
\begin{equation}
\hbar\Gamma_{\mathrm{PA}}\approx\frac{\pi I}{c\epsilon_{0}}|\expectation{d}|^{2}\hspace{1mm}F,
\end{equation}
where $I$ is the PA laser intensity, $\expectation{d}$ is the electric dipole matrix element, and $F$ is the free-to-bound Franck-Condon factor.  The total ground state formation rate coefficient, $K$, is then found by taking the thermal average of the product of the relative velocity and the photoassociation cross-section, summing over partial waves, which is given as
\begin{equation}
K=\bigg<v_{\mathrm{rel}}\frac{\pi}{k^{2}}\sum_{\ell}(2\ell+1)\frac{\Gamma_{\mathrm{PA}}\gamma_{\mathrm{nat}}}{(\Gamma_{\mathrm{PA}}+\gamma_{\mathrm{nat}})^{2}}\bigg>,
\end{equation} 
where $k$ is the de Broglie wavevector, $\gamma_{\mathrm{nat}}$ is the PA state linewidth, $v_{\mathrm{rel}}$ is the relative velocity of the collision partners, and the brackets denote a thermal average. At our temperatures, we expect only $v_{\mathrm{rel}}$ and $k$ to vary significantly over the thermal average; $\Gamma_{\mathrm{PA}}$ is taken to be constant, as the scattering state wavefunction, $\ket{\chi(R)}$, should only vary slowly over the relevant energies.

Finally, this rate coefficient is multiplied by the product of the atomic densities and integrated over the volume illuminated by the PA laser beam to give the total ground state molecular formation rate, $R$, 
\begin{equation}
R=K\int n_{\mathrm{Rb}}n_{\mathrm{Cs}}dV.
\end{equation}

Based on our experimental parameters, the calculated FCF (330 E$_{\mathrm{h}}^{-1}$, where E$_{\mathrm{h}}$ is the Hartree energy), and a crudely approximate value for the electric dipole matrix element including all angular factors ($\expectation{d}\!\sim\!2ea_{0}$), we expect a total PA rate of approximately $R\sim5\times10^{5}$ molecules/s.  This value takes into account that the PA laser beam is smaller than the atomic clouds in the MOTs. 

The ground state formation rate, $R_{X}$, is scaled by the branching fraction for decay from the PA state to the X($v$=0) state, and is given by $R_{X}=bR$. The value of $b$ is determined by a combination of two factors: the electronic branching fraction and the bound-to-bound FCF. We crudely estimate the electronic state branching fraction as follows.   The (2)$^{3}\Pi_{0^{+}}\!\to$X$^{1}\Sigma^{+}$ electric dipole transition is nominally forbidden, due to the change in total spin $S$.  However, mixing of singlet and triplet states due to the spin-orbit interaction is a significant effect in heavy molecules like RbCs, and such mixing, for example, of the (2)$^{3}\Pi_{0^{+}}$ state with the nearby (3)$^{1}\Sigma^{+}$ state, leads to a non-zero transition amplitude. Very crudely, we can expect a singlet mixing amplitude due to the spin-orbit interaction of $\sim$(Z$\alpha)^{2}\approx$(0.16) for Cs \cite{landau1977quantum}, hence a transition probability reduced by a factor of $(Z\alpha)^4\!\approx\!0.03$ compared to a fully allowed electric dipole transition.  To calculate the electronic state branching fraction, we must also take into account the scaling of the spontaneous decay rate with $\omega^{3}$, where $\omega$ is the frequency of the decay radiation.  The frequency for the $(2)^{3}\Pi_{0^{+}}\!\to$ X decay ($\sim$15700 cm$^{-1}$) is a factor of 1.3 larger than that for the fully allowed decay to the a$^{3}\Sigma^{+}$ state ($\sim$12000 cm$^{-1}$).  Hence the electronic state branching fraction for (2)$^{3}\Pi_{0^{+}}\!\to$X decay should be $\sim 1.3^{3} (Z\alpha)^4 \approx 0.06$. As shown in Figure \ref{FCF}, the ($v'$=10$\to\!v$=0) bound-to-bound FCF has the value $F\approx$0.1. Hence we finally estimate a vibronic state branching ratio, $b\sim$0.006. Putting this together with the calculated photoassociation rate, $R$, we therefore predict an approximate total ($v$=0) production rate of $\sim$3$\times10^{3}$ molecules/s under our particular experimental conditions. 

This calculated result can be compared to the approximately calibrated signal in our experiment. Based on our signal size, detector gain, crudely estimated ionization efficiency ($\sim\!5\%$) \cite{892556}, and molecule fraction in the detection region for a given REMPI pulse ($\approx$0.025, due to molecular motion at a temperature $T\approx$100 $\mu$K), we estimate the observed ($v$=0) production rate to be $\sim$6$\times10^{3}$ molecules/s, where the primary uncertainties come from the Channeltron detector gain and the ionization efficiency.  This is in qualitative agreement with our theoretical estimate.

We also briefly consider the unexpected X($v$=0) formation observed from PA through the $\Omega$=0$^{-}$ level first observed in Ref. \cite{C1CP21497G}. It appears that this PA state decays to the X$^{1}\Sigma^{+}$ state via a multi-step process (e.g. via the mixed B$^{1}\Pi_{1}$ and b$^{3}\Pi_{1}$/c$^{3}\Sigma_{1}$ states \cite{rbcsprospects}), since the $\Omega$=0$^{-}\!\nrightarrow\!\Omega$=0$^{+}$ selection rule forbids direct decay \cite{herzberg1950molecular}. Compared to the case of direct, single-photon decay, this two-photon process should be suppressed by the $\omega^{3}$ dependence of the spontaneous decay rate to the higher-lying intermediate levels. However, strong singlet/triplet mixing in the intermediate states may enhance the relative formation rate of X$^{1}\Sigma^{+}$($v$=0) molecules for this process, compared to the direct, nominally forbidden decay from the (2)$^{3}\Pi_{0^{+}}$ state. Further, the PA laser creates a potential well that is a few hundreds of $\mu$K deep for Cs atoms for this resonance ($\approx$8 cm$^{-1}$ red-detuned relative to the Cs D2 line) under our conditions. This may lead to an enhancement in the observed X$^{1}\Sigma^{+}$($v$=0) formation rate, beyond that expected simply from the PA rate and decay branching ratio.  We note that this mechanical effect of the PA laser was clearly evident on this resonance; at very high PA laser intensities for this particular resonance, the Cs MOT was visibly perturbed, reducing PA efficiency.

\section{Outlook}
Having demonstrated photoassociative formation of vibronic ground state molecules, we now discuss a potentially simple pathway to create large trapped samples of rovibronic X($v$=$J$=0) ground state molecules, by accumulating these molecules in an optical trap. We begin with a brief discussion of rotational state populations. Although rotational levels of the X($v$=0) state are unresolved with our pulsed laser detection scheme, the expected relative rotational populations can be calculated based on the associated H\"{o}nl-London factors for spontaneous decay.  For the (2)$^{3}\Pi_{0^{+}}$($J_{\mathrm{PA}}$=1) $\to$ X$^{1}\Sigma^{+}$($J_{X}$=0,2) transition, the relative line strength in emission is S$_{J=2}$/S$_{J=0}$=2. Thus 1/3 of the ($v_{X}$=0) molecules should be in the $J_{X}$=0 rotational level.  

These rovibronic ground state molecules can be continuously formed and accumulated in an optical dipole trap (as demonstrated for vibrationally-excited molecules in Refs. \cite{PhysRevLett.96.023202, hudsoninelastic2008, deiglmayr2011inelastic}). However, inelastic collisions leading to trap loss compete with molecular formation. Specifically, two-body collisions with Rb atoms and rovibronically excited molecules (RbCs$^{*}$) are predicted to cause trap loss, while collisions with Cs atoms and rovibronic ground state molecules are predicted to be purely elastic.\footnote{In this analysis we ignore collisions with electronically-excited atoms.  This seems justified if we assume the continued use of dark-SPOT MOTs, where the atomic excited state fraction is extremely small.} Thus, in order to achieve the longest possible trap lifetimes, it will be necessary to remove the rotationally and vibrationally excited molecules that will also be created via PA and subsequently trapped along with the desired X($v$=$J$=0) molecules.   

We argue that it should be possible to exploit the rapid inelastic collisions of RbCs$^{*}$ with Cs to selectively remove the unwanted molecules while retaining the rovibronic ground state molecules, as suggested in Ref. \cite{hudsoninelastic2008}. Then removing the atoms with a resonant push laser beam will leave a collisionally-stable sample of rovibronic ground state molecules. Here we discuss a scheme to accomplish this, consisting of two steps.  In the first step, RbCs molecules are formed by PA in the presence of Rb and Cs MOTs, and their population accumulates in a co-located optical trap.  Next, the Rb atoms are removed, while the Cs atoms are kept to ``scrub" the unwanted molecular states.

During the initial stage of PA in the presence of the optical trap and atomic MOTs, the governing rate equations for molecular creation and destruction are given by
\begin{eqnarray}
\label{ground}\dot{n}{_{\mathrm{RbCs}}}  =&  \tilde{b}K_{\mathrm{PA}}n_{\mathrm{Rb}}n_{\mathrm{Cs}}-K_{(\mathrm{RbCs-Rb})}n_{\mathrm{RbCs}}n_{\mathrm{Rb}}\\
&-K_{(\mathrm{RbCs-RbCs^{*}})}n_{\mathrm{RbCs}}n_{\mathrm{RbCs^{*}}}, \nonumber 
\\
\label{excited}\dot{n}{_{\mathrm{RbCs^{*}}}}  =&  (1-\tilde{b})K_{\mathrm{PA}}n_{\mathrm{Rb}}n_{\mathrm{Cs}}-K_{(\mathrm{RbCs^{*}-Rb})}n_{\mathrm{RbCs^{*}}}n_{\mathrm{Rb}} \\
&-K_{(\mathrm{RbCs^{*}-Cs})}n_{\mathrm{RbCs^{*}}}n_{\mathrm{Cs}} -K_{(\mathrm{RbCs^{*}-RbCs})}n_{\mathrm{RbCs^{*}}}n_{\mathrm{RbCs}}\nonumber\\
&-K_{(\mathrm{RbCs^{*}-RbCs^{*}})}n_{\mathrm{RbCs^{*}}}^{2}, \nonumber
\end{eqnarray}
where $\tilde{b}$=$b$/3 is the branching ratio to the $v$=$J$=0 level, $n_{x}$ is the density of species $x$, $K_{PA}$ is the rate coefficient for photoassociation, and $K_{(x-y)}$ is the inelastic collision rate coefficient for species $x$ and $y$. Here $x$=RbCs$^{*}$ refers to RbCs molecules in any state except X($v$=$J$=0) state, and $x$=RbCs refers only to rovibronic ground state molecules. We conservatively assume that all collisions proceed at the unitarity-limited rate ($K_{\mathrm{max}}=\expectation{v_{\mathrm{rel}}\frac{\pi}{k^{2}}}\approx3\times10^{-11}$ cm$^{3}$/s at 100 $\mu$K), as suggested by previous work in Refs. \cite{PhysRevLett.96.023202,Ospelkaus12022010,hudsoninelastic2008,deiglmayr2011inelastic}. Further, we set $K_{\mathrm{PA}}$ to the same value, as it should be possible to saturate the PA rate with sufficient laser power \cite{PhysRevLett.91.080402}. Then in the limit of small branching, $\tilde{b}\ll 1$ ($\tilde{b}\sim$0.2\% for X($v$=$J$=0), as calculated above), and constant and equal Rb and Cs density ($n_{\mathrm{Rb}}\!=\!n_{\mathrm{Cs}}$), the steady-state density of rovibronic ground state molecules is given approximately by 
\begin{equation}
n_{\mathrm{RbCs}}\approx \frac{\tilde{b}}{\sqrt{2}}n_{\mathrm{Cs}}.
\end{equation}
For improved atomic densities of $n_{\mathrm{Rb}}$=$n_{\mathrm{Cs}}$=5$\times$10$^{11}$ cm$^{-3}$, as achieved in Ref. \cite{hudsoninelastic2008}, and  including the effects of atomic loss from the MOTs due to PA, the ground state molecular density maximizes at $n_{\mathrm{RbCs}}\approx$3$\times$10$^{8}$ cm$^{-3}$ after $\approx$100 ms for $\tilde{b}$=$0.002$.

\begin{figure}
\begin{center}
\includegraphics[width=0.75\columnwidth]{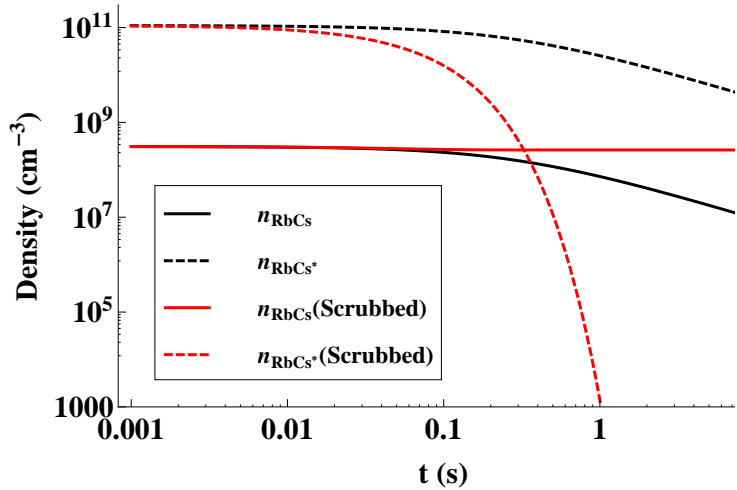}
\end{center}
\caption{\label{scrub}Simulated population dynamics of excited (dashed lines) and ground state (solid lines) molecular densities during the ``scrubbing" stage. Initial densities are given for PA with $n_{\mathrm{Rb}}$=$n_{\mathrm{Cs}}$=5$\times10^{11}$ cm$^{-3}$. Black lines show dynamics in the absence of co-trapped Cs atoms. Red lines show dynamics with co-trapped Cs density of 6$\times$10$^{11}$cm$^{-3}$. All inelastic rate coefficients are set to K$_{\mathrm{max}}=3\times10^{-11}$ cm$^{3}$/s.}
\end{figure}

Once the maximum density is achieved, the Rb atoms can be rapidly removed from the trap by a resonant push beam.  Then the trapped sample contains the desired RbCs molecules, as well as Cs atoms and undesired RbCs$^{*}$ states.  The population dynamics in this stage of the experiment are described by the rate equations
\begin{eqnarray}
\dot{n}{_{\mathrm{RbCs}}}=&-K_{(\mathrm{RbCs-RbCs^{*}})}n_{\mathrm{RbCs}}n_{\mathrm{RbCs^{*}}}, \\
\dot{n}{_{\mathrm{RbCs^{*}}}}=&-K_{(\mathrm{RbCs^{*}-Cs})}n_{\mathrm{RbCs^{*}}}\tilde{n}_{\mathrm{Cs}} -K_{(\mathrm{RbCs^{*}
-RbCs})}n_{\mathrm{RbCs^{*}}}n_{\mathrm{RbCs}}\nonumber\\
&-K_{(\mathrm{RbCs^{*}-RbCs^{*}})}n_{\mathrm{RbCs^{*}}}^2, 
\end{eqnarray}
where $\tilde{n}_{\mathrm{Cs}}$ is the density of Cs atoms in the trap. Using the calculated molecular densities after PA as the initial conditions, these rate equations can be solved numerically. As shown in Figure \ref{scrub}, for even modest co-trapped atomic Cs densities, excited molecules can be removed quickly with only a small reduction of the ground state molecular density. For a co-trapped Cs density of $\tilde{n}_{\mathrm{Cs}}$=6$\times10^{11}$ cm$^{-3}$ (as in e.g. Ref. \cite{hudsoninelastic2008}) and $\tilde{b}$=$0.002$, $n_{\mathrm{RbCs}}$ stabilizes at 3$\times10^{8}$ cm$^{-3}$ after $\approx$300 ms, a 15$\%$ reduction from the initial ground state molecular density (just before removal of the Rb atoms).

Now we consider realistic experimental conditions and the potential for this approach to create a sample of trapped X($v$=$J$=0) molecules. For a trap radius of 200 $\mu$m (as in e.g. Ref. \cite{hudsoninelastic2008}), reaching saturation of the PA transition using our experimental fit (calculated) value requires a laser power of approximately 5(16) W.  Although it is difficult to achieve this power with a single laser beam pass, a relatively low finesse ($\mathcal{F}\!\sim\!5\!-\!20$) power build-up cavity can be used to increase the intensity to the desired level over the trap volume. We assume this saturation intensity can be achieved over the entire volume. The final results of this discussion can be summarized as follows, using what we consider to be realistic experimental parameters:  for initial MOT densities of $n_{\mathrm{Rb}}$=$n_{\mathrm{Cs}}$=5$\times10^{11}$ cm$^{-3}$, an X($v$=$J$=0) branching ratio of $\tilde{b}$=$0.002$, a co-trapped Cs density of $\tilde{n}_{\mathrm{Cs}}$=6$\times$10$^{11}$ cm$^{-3}$, and a trap volume of $\pi$(200 $\mu$m)$^{2}\times$1 mm, we calculate a X($v$=$J$=0) molecular density (number) $n_{\mathrm{RbCs}}$=3$\times10^{8}$ cm$^{-3}$ ($N_{\mathrm{RbCs}}$=4$\times$10$^{4}$) after approximately 400 ms (100 ms for loading, and 300 ms for ``scrubbing" of unwanted states).  This seems quite promising as a simple method for producing a sample of ultracold, polar RbCs molecules in their rovibronic ground state.

\section{Conclusion}
We have demonstrated numerous pathways to the production of rovibronic ground state of $^{85}$RbCs molecules, through short-range photoassociation to the (2)$^{3}\Pi_{0^{+}}$ state followed by radiative decay to the X($v$=$J$=0) state. We have predicted and experimentally verified which of these PA levels most efficiently create X$^{1}\Sigma^{+}$($v$=0) molecules. With this straightforward path to producing molecules in the rovibronic ground state, it may be possible to accumulate a large sample of molecules in an optical trap. We have calculated the expected density of ground state molecules, and proposed a mechanism to purify the sample of unwanted species.  We believe this represents a promising and extremely simple route to producing trapped samples of ultracold polar molecules.  We note as well that while the particular PA pathway used here is convenient in many ways, there may be more promising pathways in RbCs. For example, the strongly mixed (1)$^{3}\Pi_{0^{+}}$/(2)$^{1}\Sigma^{+}$ complex of states \cite{C1CP21769K}, or states with resonant coupling, as seen in KRb in Ref. \cite{PhysRevA.86.053428}, may provide the required large transition strengths to both the atomic scattering state and the rovibrational ground state.

\ack{We acknowledge fruitful discussions with J.T. Kim and W.C. Stwalley, especially for pointing out the work done in Ref. \cite{doi:10.1021/jp803360w}.  We also gratefully acknowledge funding from the DOE, NSF, and AFOSR-MURI.}

\section*{References}
\bibliographystyle{iopartnum}
\bibliography{PAPaperv3}

\end{document}